\documentclass[usenatbib,usegraphicx,useAMS,usenatbib]{mn2e}

\newcommand{\hii}{{H{\scriptsize II} }}

\newcommand{\choh}{CH$_3$OH}

\newcommand{\hcccn}{HC$_3$N}

\newcommand{\kms}{km\,s$^{-1}$}

\title[ATCA Observations of G305.2+0.2]{ATCA 1.2\,cm Observations of the Massive Star Forming Region G305.2+0.2}
\author[A. J. Walsh et al.]{Andrew J. Walsh$^{1,2}$\thanks{E-mail:
andrew.walsh@jcu.edu.au}, Jacqueline F. Chapman$^3$,
\newauthor
Michael G. Burton$^{1}$, Mark Wardle$^3$ and T. J. Millar$^4$\\
$^{1}$School of Physics, University of New South Wales, Sydney, NSW 2052, Australia\\
$^{2}$School of Maths, Physics and IT, James Cook University, 4814, Australia\\
$^{3}$Department of Physics, Macquarie University, Sydney, NSW 2109, Australia\\
$^{4}$Astrophysics Research Centre, School of Mathematics and Physics, Queen's University Belfast,
Belfast BT7 1NN, UK}
\begin{document}



\maketitle

\label{firstpage}

\begin{abstract}
We report on Australia Telescope
observations of the massive star forming region G305.2+0.2 at 1.2\,cm. We detected emission in
five molecules towards G305A, confirming its hot core nature. We determined a rotational temperature of 26\,K
for methanol. A non-LTE excitation calculation suggests a kinematic temperature of order 200\,K.
A time dependent chemical model is also used to model the gas phase
chemistry of the hot core associated with G305A. A comparison with the
observations suggest an age of between 2 $\times 10^4$ and 1.5 $\times 10^5$ years.
We also report on a
feature to the SE of G305A which may show weak Class I methanol maser
emission in the line at 24.933\,GHz.
The more evolved source G305B does not show emission in any of the line tracers,
but strong Class I methanol maser emission  at 24.933\,GHz is found 3\arcsec~to the east.
Radio continuum emission at 18.496\,GHz is detected towards two \hii regions. The implications of
the non-detection of radio continuum emission toward G305A and G305B are also discussed.
\end{abstract}


\section{Introduction}
The study of the early stages of massive star formation (MSF) in our Galaxy has, until recently,
received relatively little attention compared to low mass star
formation. However, with the advent of telescopes,
particularly in the millimetre (mm) and sub-mm part of the electromagnetic
spectrum, it has become possible to study the earliest stages of MSF.
This is the strongest part of their spectral energy distribution
(SED) measureable from the ground. It also contains a myriad of
complex molecular transitions, allowing investigation of the
physics and chemistry surrounding these young sources.

G305.2+0.2 is a site of MSF in the southern Galactic plane. Class II methanol
masers were reported in two positions by \citet{norris93}:
G305.21+0.21 (G305A) and G305.20+0.21 (G305B), separated by
approximately 22\arcsec.
\citet{walsh99} observed this region in the near-infrared and found that whilst
G305A is not associated with any infrared source, G305B is coincident with a
bright and very reddened infrared source. More sensitive near-infrared observations
by \citet{debuizer03} indicate that there is indeed a weak infrared source
coincident with G305A, however, it is not clear if it is
associated with the maser site or is instead an unrelated foreground star.
Observations in the mid-infrared (10.5
and 20\,$\mu$m) by \citet{walsh01} confirm the infrared source associated with
G305B has a steeply rising spectral energy distribution towards longer wavelengths.
However, no mid-infrared source was found associated with G305A.
The Galactic Legacy Infrared Mid-Plane Survey Extraordinaire
(GLIMPSE) is a legacy science program of the {\it Spitzer Space Telescope}. It confirms
the detection of a weak infrared source at the position of G305A at 3.6 and 4.5$\mu$m. However, the region
shows much extended emission (see Figures 1a and 3b), which makes photometry unreliable.

\citet{hill05} have detected a strong 1.2\,mm continuum source whose peak coincides
with G305A (see figure 1 of \citet{walsh06}).
The 1.2\,mm map also shows some extended emission in the direction of G305B, but the
extended emission was poorly resolved. Because G305B appears to dominate the luminosity at
shorter wavelengths (up to about 20\,$\mu$m) and G305A appears to dominate at longer
wavelengths (1.2\,mm), there will be a transition point at far-infrared wavelengths
where the two sources will have equal
brightnesses. Thus, it is impossible to tell which source is the most luminous, based on
infrared data alone.
Neither maser site is coincident with any detected radio continuum source \citep{phillips98}.
However, one might be expected from G305B, since extrapolation of its SED
\citep{walsh01} indicates it is powered by a star bright enough to produce
an observable ultracompact (UC) \hii region.
%
%
Furthermore, it is not known why the brightest maser site, G305A is not associated
with a bright infrared source, nor a radio continuum source and yet appears to be a strong
source of continuum emission at 1.2\,mm.

\citet{walsh06} found G305A to be the local peak of emission for 1--0 transitions of
$^{13}$CO, HCO$^+$ and N$_2$H$^+$, as well as transitions of CH$_3$CN and CH$_3$OH.
NH$_3$ (1,1), (2,2) and (4,4) emission  has been detected coincident with G305A
(Steven Longmore, private communication 2007).
This suggests G305A is a hot core and may be the most massive component in the region.
The line and dust continuum emission appear to be slightly elongated in the direction of
G305B. Therefore it appears G305B may be associated with a secondary hot core,
which was spatially unresolved with the previous observations.
\citet{walsh06} also identified the possible prestellar massive core G305SW, which was only detected 
in N$_2$H$^+$ and shows narrow lines, compared to the emission elsewhere in the field of view.
In this paper we attempt to better understand the nature of these sources by investigating both
molecular line and continuum emission at 1.2\,cm.

\section{Observations}
Observations towards G305.2+0.2 were carried out using two configurations of the
ATCA in 2005. The first observations took place on 2005, May, 3-4th and utilised the
750A configuration. The second observations took place on 2005, September, 4th and utilised
the H214 configuration. Details of each observing session are given below.

\subsection{2005, May observations}
Using the 750A configuration, we observed G305 with baselines ranging from 77 to 735\,m,
with five antennas, and baselines up to 3750\,m when including antenna 6. However, data
from antenna 6 were only used when observing masers. The primary beam for these observations
was between 2.0\arcmin~and 2.6\arcmin, and the synthesised beam was between
3.2\arcsec~and 4.9\arcsec, when only 5 antennas were used. Observations were made towards a
single position, with pointing centre 13$^h$ 11$^m$ 10.6$^s$, right ascension and
$-$62$^\circ$ 34$^\prime$ 38$^{\prime\prime}$ declination (J2000). We spent a total
of 12 hours observing G305 in this configuration, which included time for primary calibration
on 1934-638, bandpass calibration on 1253-055 and phase calibration on 1414-59. We observed in
a frequency-switching mode, with main (spectral line) frequencies being 18199\,MHz and 24932\,MHz,
which include emission lines of cyanoacetylene (\hcccn~(2--1)) and the K=2 ladder of methanol
(\choh), respectively. Assumed rest frequencies for these lines are given in Table \ref{restfreq}.
Each frequency was observed with 256 channels over a bandwidth of 16\,MHz, resulting in a velocity
resolution of 1.0\,\kms~and 0.75\,\kms~per channel at the \hcccn~and \choh~line frequencies,
respectively. In addition to the spectral lines, continuum emission was observed at 18496\,MHz
with a bandwidth of 128\,MHz.

\subsection{2005, September observations}
Using the H214C hybrid configuration, we observed G305 with baselines ranging from 82\,m to 240\,m
with five antennas. The primary beam for these observations
was between 2.0\arcmin~and 2.6\arcmin, and the synthesised beam was between
8.6\arcsec~and 11.8\arcsec. Observations were made with a
single point, with pointing centre 13$^h$ 11$^m$ 6.25$^s$, right ascension and
-62$^\circ$ 34$^\prime$ 53$^{\prime\prime}$ declination (J2000). The reason for the different pointing
centre to the 2005, May observations was to cover the newly discovered source G305SW \citep{walsh06}.
We spent a total of 11 hours observing G305 in this configuration, which included time for primary calibration
on 1934-638, bandpass calibration on 0537-441 and phase calibration on 1414-59. We observed in
a frequency-switching mode, with five pairs of spectral line frequencies, the details of which
are given in Table \ref{restfreq}.

Observations made in this session have poorer spatial and spectral resolution than the 2005,
May session. The main implication of this is that the spectral resolution is insufficient to
identify most systemic motions, although we are able to identify detections and discriminate between
emission from G305A and G305B.

\begin{table*}
\caption{\protect\footnotesize{Properties of observed lines. The notation for each line in the second column
either conforms to the structure of (J$_{\rm upper}$ -- J$_{\rm lower}$)
or (J$_{-{\rm K},+{\rm K}}$ -- J$_{-{\rm K},+{\rm K}}$).}}
\label{restfreq}
\begin{center}
\begin{tabular}{ccccccc}
\hline
Date & Observed & Rest & Primary & Synthesized & Bandwidth & Velocity\\
& Line & Frequency & Beam & Beam & (MHz) & Resolution\\
&& (MHz) & (\arcmin) & (\arcsec) && (\kms)\\
\hline
2005 MAY 3-4 & HC$_3$N (2--1) & 18196.226 & 2.6 & 4.9 & 16 & 1.0\\
&&&&&&\\
2005 MAY 3-4 & CH$_3$OH (3$_{2,1}$ -- 3$_{1,2}$) E & 24928.700 & 2.0 & 3.2 & 16 & 0.75\\
             & CH$_3$OH (4$_{2,2}$ -- 4$_{1,3}$) E & 24933.470 &     &     &    &     \\
             & CH$_3$OH (2$_{2,0}$ -- 2$_{1,1}$) E & 24934.380 &     &     &    &     \\
&&&&&&\\
2005 SEP 4 & NH$_3$ (1,1)                          & 23694.496 & 2.0 & 9.0 & 32 & 3.2\\
           & NH$_3$ (2,2)                          & 23722.633 & 2.0 & 9.0 & 32 & 3.2\\
&&&&&&\\
2005 SEP 4 & H$_2$O (6$_{1,6}$ -- 5$_{2,3}$)       & 22235.080 & 2.1 & 9.6 & 32 & 3.4\\
           & CH$_3$OCHO (2$_{1,2}$ -- 1$_{1,1}$) E & 22827.741 & 2.1 & 9.4 & 32 & 3.3\\
           & CH$_3$OCHO (2$_{1,2}$ -- 1$_{1,1}$) A & 22828.134 & 2.1 & 9.4 & 32 & 3.3\\
           & NH$_3$ (3,2)                          & 22834.185 & 2.1 & 9.4 & 32 & 3.3\\
&&&&&&\\
2005 SEP 4 & CH$_3$OH (8$_{2,6}$ -- 8$_{1,7}$) E   & 25294.417 & 1.9 & 8.6 & 64 & 11.9\\
           & DC$_3$N (3--2)                        & 25329.437 & 1.9 & 8.5 & 64 & 12.0\\
&&&&&&\\
2005 SEP 4 & OCS (2--1)                            & 24325.927 & 2.0 & 8.8 & 64 & 12.3\\
&&&&&&\\
2005 SEP 4 & CH$_3$CN (1--0)                       & 18399.892 & 2.6 & 11.6 & 32 & 4.1\\
&&&&&&\\
2005 SEP 4 & DC$_3$N (2--1)                        & 16886.310 & 2.8 & 12.7 & 64 & 17.8\\
           & H91$\beta$ (RRL)                      & 16894.200 & 2.8 & 12.7 & 64 & 17.8\\
           & HC$_7$N (15--14)                      & 16919.979 & 2.8 & 12.7 & 64 & 17.7\\
           & HC$_5$N (7--6)                        & 18638.616 & 2.6 & 11.5 & 32 & 4.0\\
\hline
\end{tabular}
\end{center}
\end{table*}

\section{Results}
Approximately half the lines that were searched were detected, and details of these lines
are given in Table \ref{nondets}. The table also includes details of those lines that were not detected,
together with their upper limits. Since the 2005 September observing run was conducted using a relatively
poor spatial and spectral resolution, we do not know line widths or precise locations for most of the lines
detected during this time. In the last column of Table \ref{nondets}, we list the position
of the emission and nearest association if it is not found coincident with G305A or G305B.
The observations of
methanol and cyanoacetylene conducted in 2005 May have both higher spatial and spectral resolution, and so
we report more detailed results on each below.

\begin{table*}
\caption{\protect\footnotesize{Properties of detections and non-detections.
The notation for each line either conforms to the structure of
(J$_{\rm upper}$ -- J$_{\rm lower}$)
or (J$_{-{\rm K},+{\rm K}}$ -- J$_{-{\rm K},+{\rm K}}$). Only detections for each transition are reported, except
where no detection anywhere was made. In the case of a non-detection, then 1$\sigma$ upper limit for the integrated
intensity is given. The last column lists names that are used in the text as a reference for the position of the
emission.}}
\label{nondets}
\begin{center}
\begin{tabular}{cccccc}
\hline
Observed                            &         Position        & Integrated &   Peak   &  Line   & Association \\
Line                                &         (J2000)         &  Intensity & Velocity &  Width  &             \\
                                    &                         &  (K.\kms)  &  (\kms)  &  (\kms) &             \\
\hline
HC$_3$N (2--1)                      & 13 11 13.71 -62 34 40.5 &     9.6    &  -41.9   &    5.1  & G305A \\
CH$_3$OH (3$_{2,1}$ -- 3$_{1,2}$) E & 13 11 13.71 -62 34 41.0 &    12.2    &  -39.8   &    5.1  & G305A \\
CH$_3$OH (4$_{2,2}$ -- 4$_{1,3}$) E & 13 11 13.71 -62 34 41.0 &    11.8    &  -39.8   &    4.8  & G305A \\
CH$_3$OH (2$_{2,0}$ -- 2$_{1,1}$) E & 13 11 13.71 -62 34 41.0 &     9.0    &  -40.0   &    5.9  & G305A \\

CH$_3$OH (3$_{2,1}$ -- 3$_{1,2}$) E & 13 11 14.00 -62 34 46.5 &     4.0    &  -39.7   &    2.5  & G305A(SE) \\
CH$_3$OH (4$_{2,2}$ -- 4$_{1,3}$) E & 13 11 14.00 -62 34 46.5 &     5.3    &  -39.7   &    2.1  & G305A(SE) \\
CH$_3$OH (2$_{2,0}$ -- 2$_{1,1}$) E & 13 11 14.00 -62 34 46.5 &     2.3    &  -39.8   &    2.9  & G305A(SE) \\

CH$_3$OH (3$_{2,1}$ -- 3$_{1,2}$) E &             -           &  $<$1.2    &    -     &    -    & - \\
CH$_3$OH (4$_{2,2}$ -- 4$_{1,3}$) E & 13 11 10.89 -62 34 39.0 &    46      &  -42.1   & $<$0.75 & 3\arcsec~E of G305B \\
CH$_3$OH (2$_{2,0}$ -- 2$_{1,1}$) E &             -           &  $<$1.2    &    -     &    -    & - \\

NH$_3$ (1,1)                        & 13 11 13.71 -62 34 41.0 &    84      &    -     &    -    & G305A \\
NH$_3$ (2,2)                        & 13 11 13.71 -62 34 41.0 &    11      &    -     &    -    & G305A \\

OCS (2--1)                          & 13 11 14.08 -62 34 39.0 &     3.9    &  -42     &$<$12    & G305A \\
NH$_3$ (3,2)                        & 13 11 13.71 -62 34 41.0 &    15.0    &  -39     &$<$13    & G305A \\
H$_2$O (6$_{1,6}$ -- 5$_{2,3}$)     & 13 11 13.71 -62 34 41.0 & 18000      &  -40     & $<$3.4  & G305A \\
H$_2$O (6$_{1,6}$ -- 5$_{2,3}$)     & 13 11 13.50 -62 34 31.0 &    25      &  -52     & $<$3.4  & 10\arcsec~N of G305A \\
H$_2$O (6$_{1,6}$ -- 5$_{2,3}$)     & 13 11 08.29 -62 34 45.0 &     6.3    &   +4     & $<$3.4  & 14\arcsec~SW  of G305B \\
CH$_3$OCHO (2$_{1,2}$-1$_{1,1}$) E  &             -           &  $<$0.11   &    -     &    -    &  - \\
CH$_3$OCHO (2$_{1,2}$-1$_{1,1}$) A  &             -           &  $<$0.11   &    -     &    -    &  - \\
CH$_3$OH (8$_{2,6}$ -- 8$_{1,7}$) E & 13 11 10.89 -62 34 39.0 &   130      &  -43     &$<$12    &  3\arcsec~E   of G305B \\
DC$_3$N (3--2)                      &             -           &  $<$0.03   &    -     &    -    & - \\
CH$_3$CN (1--0)                     &             -           &  $<$0.05   &    -     &    -    & - \\
DC$_3$N (2--1)                      &             -           &  $<$0.03   &    -     &    -    & - \\
H91$\beta$ (RRL)                    & 13 11 08.87 -62 34 41.0 &     3.7    &  -43     &$<$18    & G305HII \\
HC$_7$N (15--14)                    &             -           &  $<$0.03   &    -     &    -    & - \\
HC$_5$N (7--6)                      &             -           &  $<$0.11   &    -     &    -    & - \\
\hline
\end{tabular}
\end{center}
\end{table*}

\subsection{\choh~(methanol)}
\begin{figure*}
\begin{tabular}{cc}
\includegraphics[width=1.0\columnwidth]{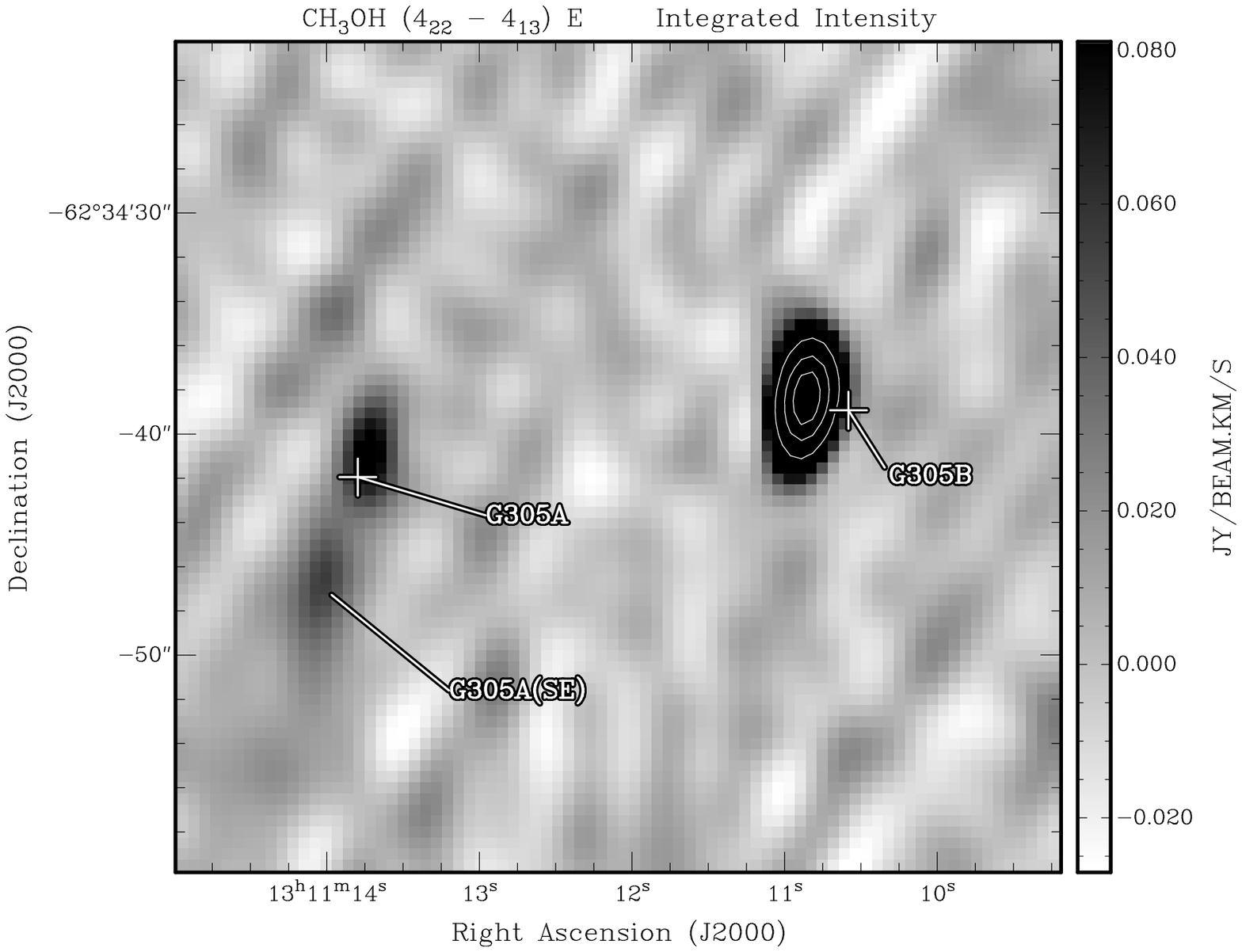}&\includegraphics[width=0.8\columnwidth]{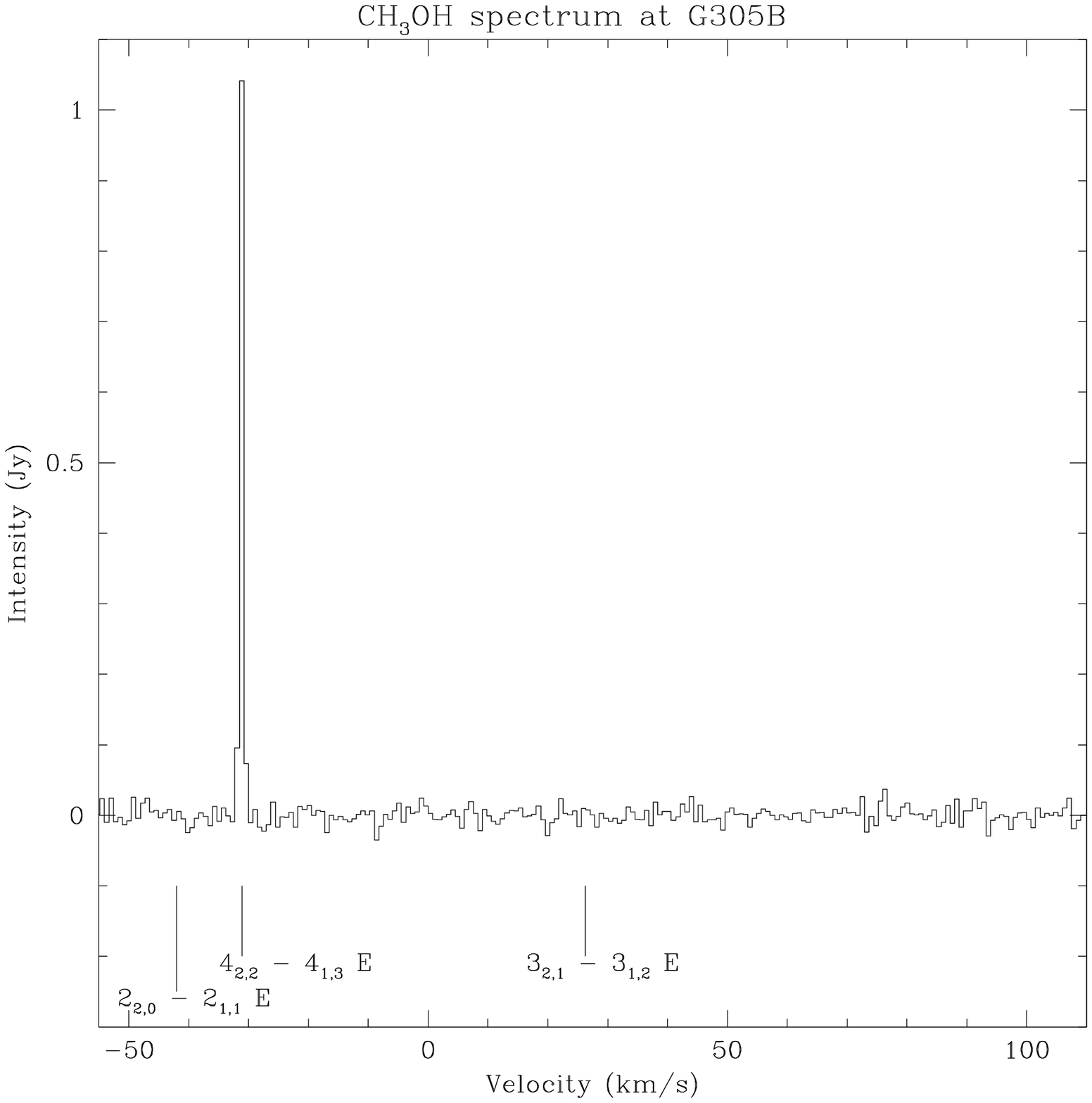}\\
\includegraphics[width=0.8\columnwidth]{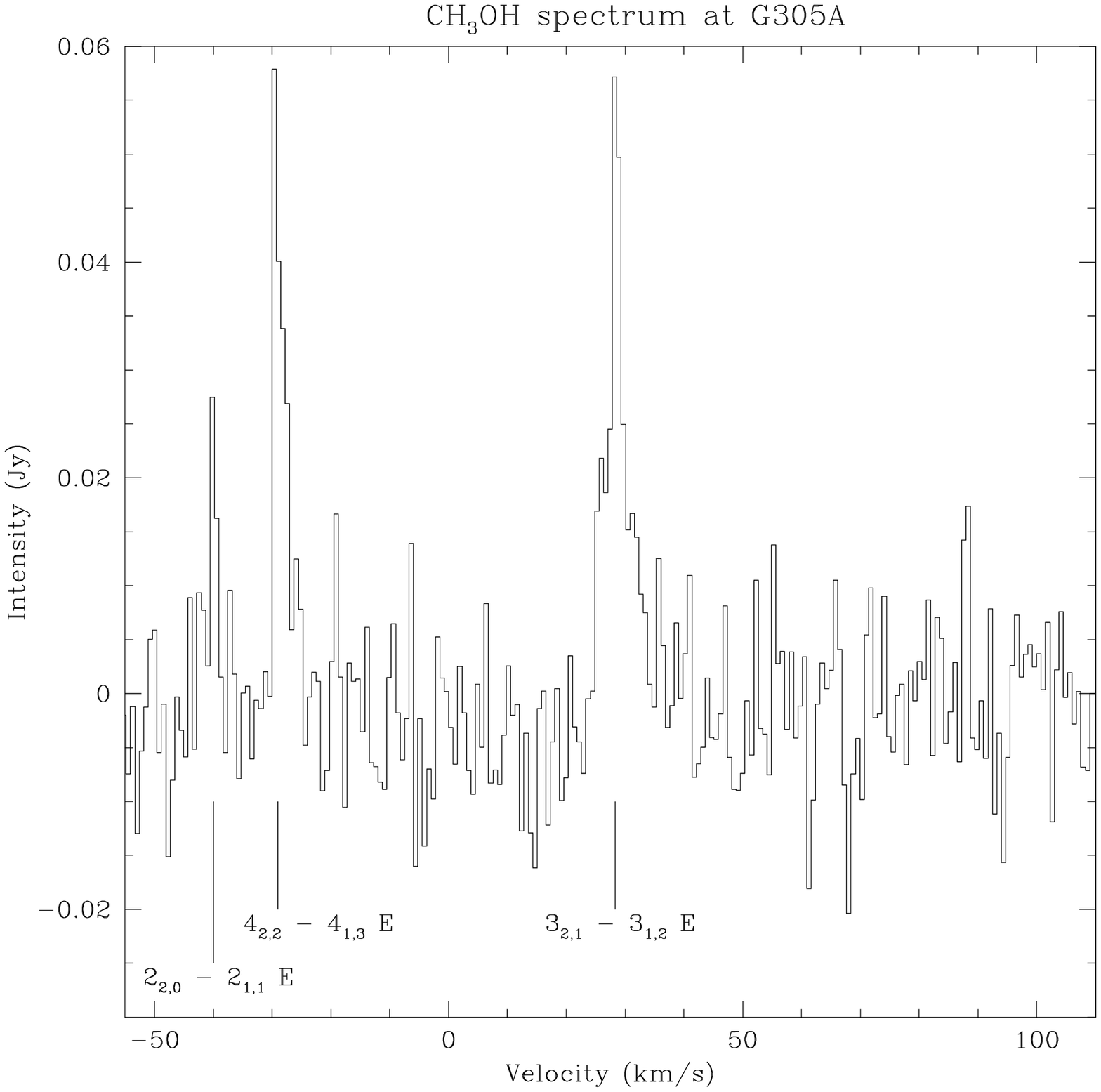}&\includegraphics[width=0.8\columnwidth]{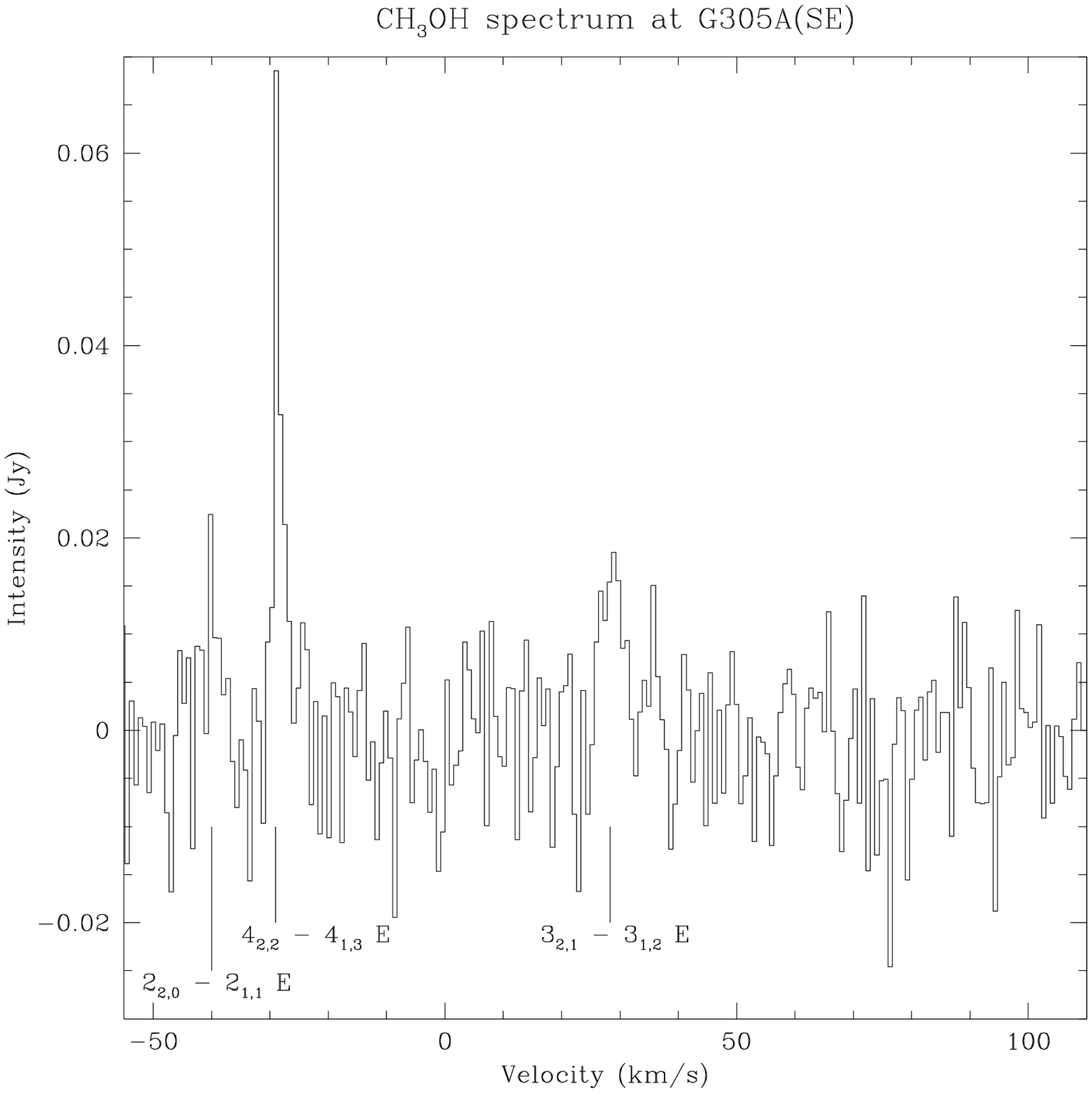}\\
\end{tabular}
\caption{{\bf (a)} \choh~integrated intensity map of the J=4 line.
Plus symbols show the positions of the 6.7\,GHz \choh~maser sites G305A and G305B.
Contour levels are 0.2, 0.4 and 0.6Jy/beam.km/s.
{\bf (b)} \choh~spectrum towards the bright line 3\arcsec~to the east of G305B, at 13 11 10.89 -62 34 38.5 (J2000).
{\bf (c)} \choh~spectrum towards G305A.
{\bf (d)} \choh~spectrum towards G305A(SE), at 13 11 14.00 -62 34 46.5 (J2000).}
\label{methimg}
\end{figure*}

The bandpass covered three lines of methanol during these observations, as listed in Table \ref{restfreq}.
We find strong, narrow line emission from the J=4 line close to the position of G305B, but not coincident
(Figure \ref{methimg}).
The emission appears to be offset from G305B by approximately 3\arcsec~to the east
(13 11 10.89 -62 34 38.5 J2000). In addition to this, weaker emission is seen in all three methanol lines
at the position of G305A (Figure \ref{methimg}). The methanol emission around G305A also shows extended
emission to the south-south-east, extending by approximately 5\arcsec . This extension, which we
refer to as G305A(SE), is strongest in
the J=4 methanol line, but the other two lines are also detected here. We note also
that the J=4 methanol line appears to be very narrow at this position. So the relative
methanol line intensities and line width are reminiscent of a weaker
version of the emission seen close to G305B.

\subsection{\hcccn~(cyanoacetylene)}
\begin{figure*}
\begin{tabular}{cc}
\includegraphics[width=1.0\columnwidth]{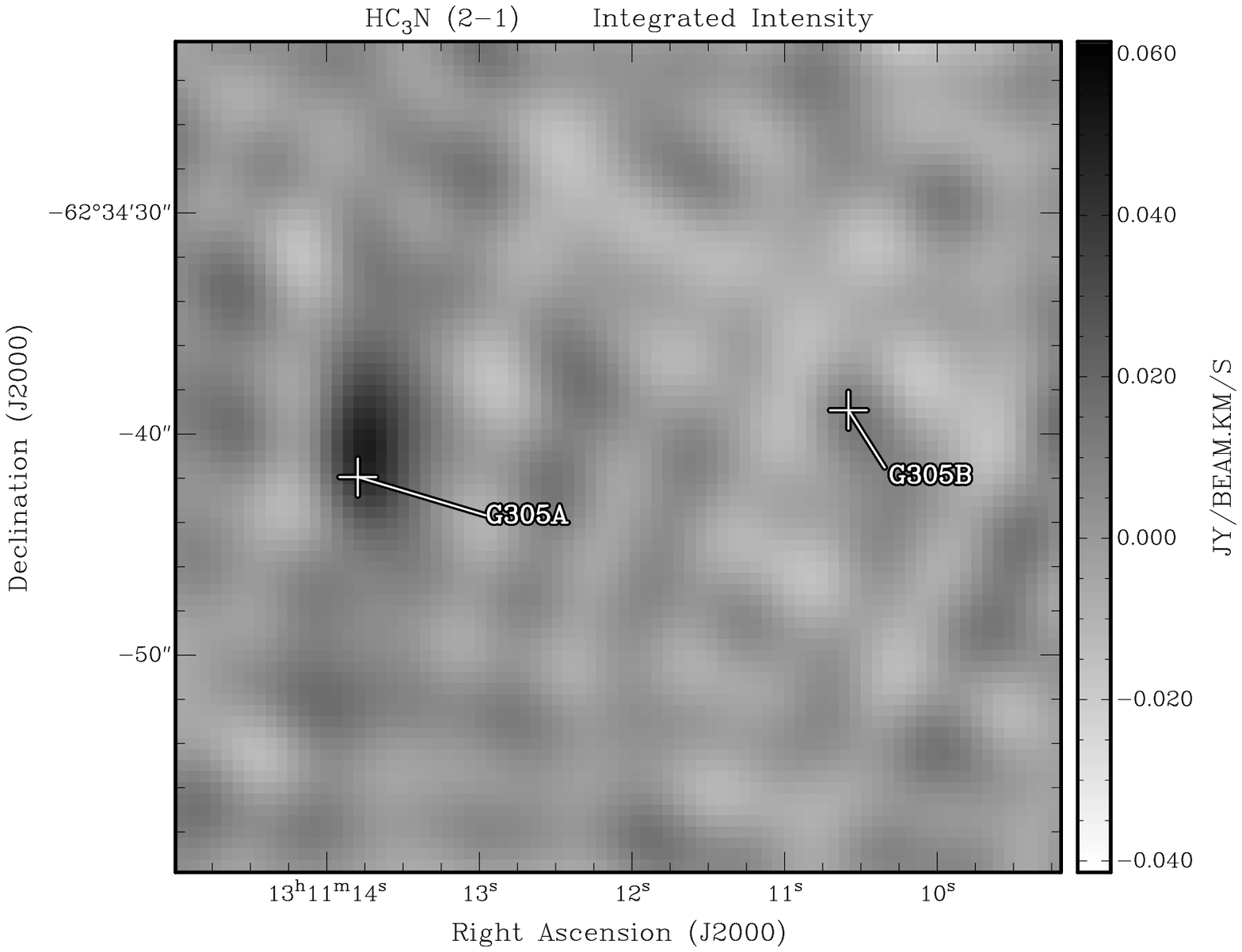}&\includegraphics[width=1.0\columnwidth]{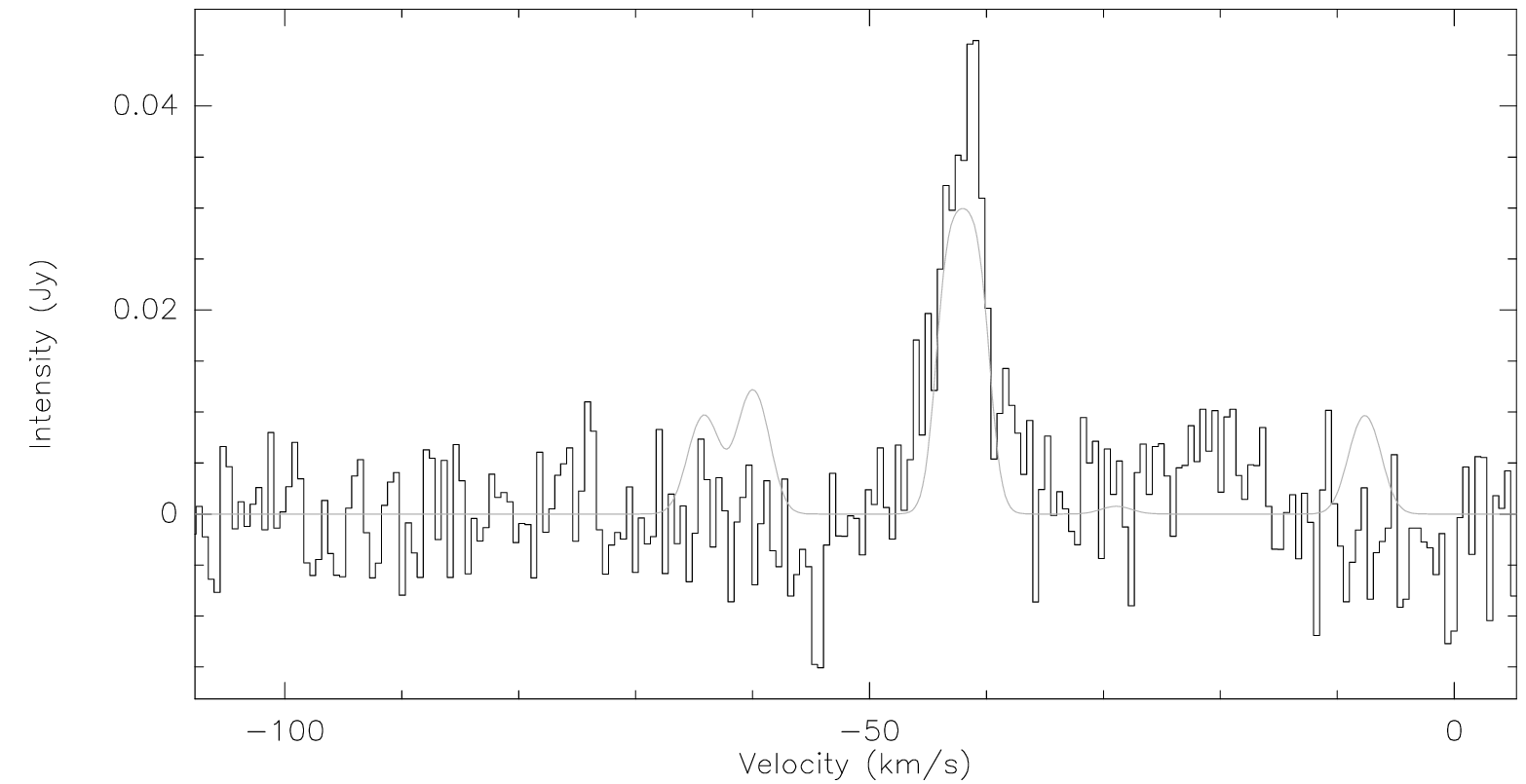}\\
\end{tabular}
\caption{{\bf (a)} \hcccn~integrated intensity map.
Plus symbols show the positions of the maser sites G305A and G305B.
{\bf (b)} \hcccn~spectrum towards G305A. The dotted line is a best fit model to the \hcccn~hyperfine structure,
holding the optical depth constant at 0.1. Because we do not see any signs of the hyperfine components, we will
assume the emission is optically thin.}
\label{hc3nimg}
\end{figure*}

We only detect cyanoacetylene coincident with the position of G305A (Figure \ref{hc3nimg}).
The emission appears to be unresolved
spatially, but is resolved spectrally with a line width of 5.1\kms . Cyanoacetylene
can be formed by neutral--neutral reactions in interstellar space \citep{fukuzawa97}, and so is likely a
precursor molecule that is present at the onset of star formation. \citet{Mea97} indicate that cyanoacetylene
can also be formed in hot cores. Their models also suggest that the cyanoacetylene abundance is greatly reduced
after about $10^5$ years. Implications of the time-dependant chemistry are discussed further in \S\ref{chemistrysection}.

\subsection{Continuum emission}
\begin{figure}
\includegraphics[width=1.0\columnwidth]{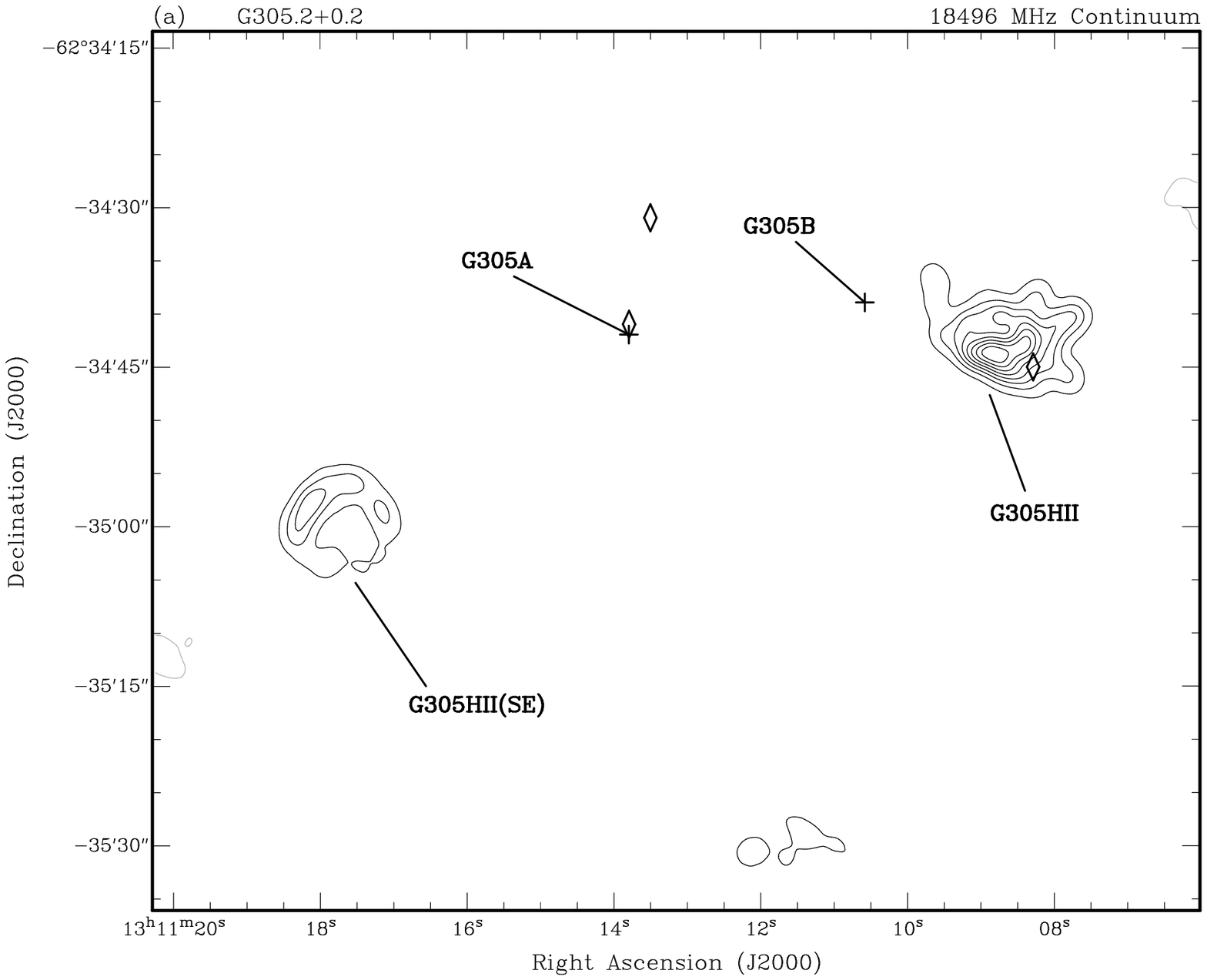}
\includegraphics[width=1.0\columnwidth]{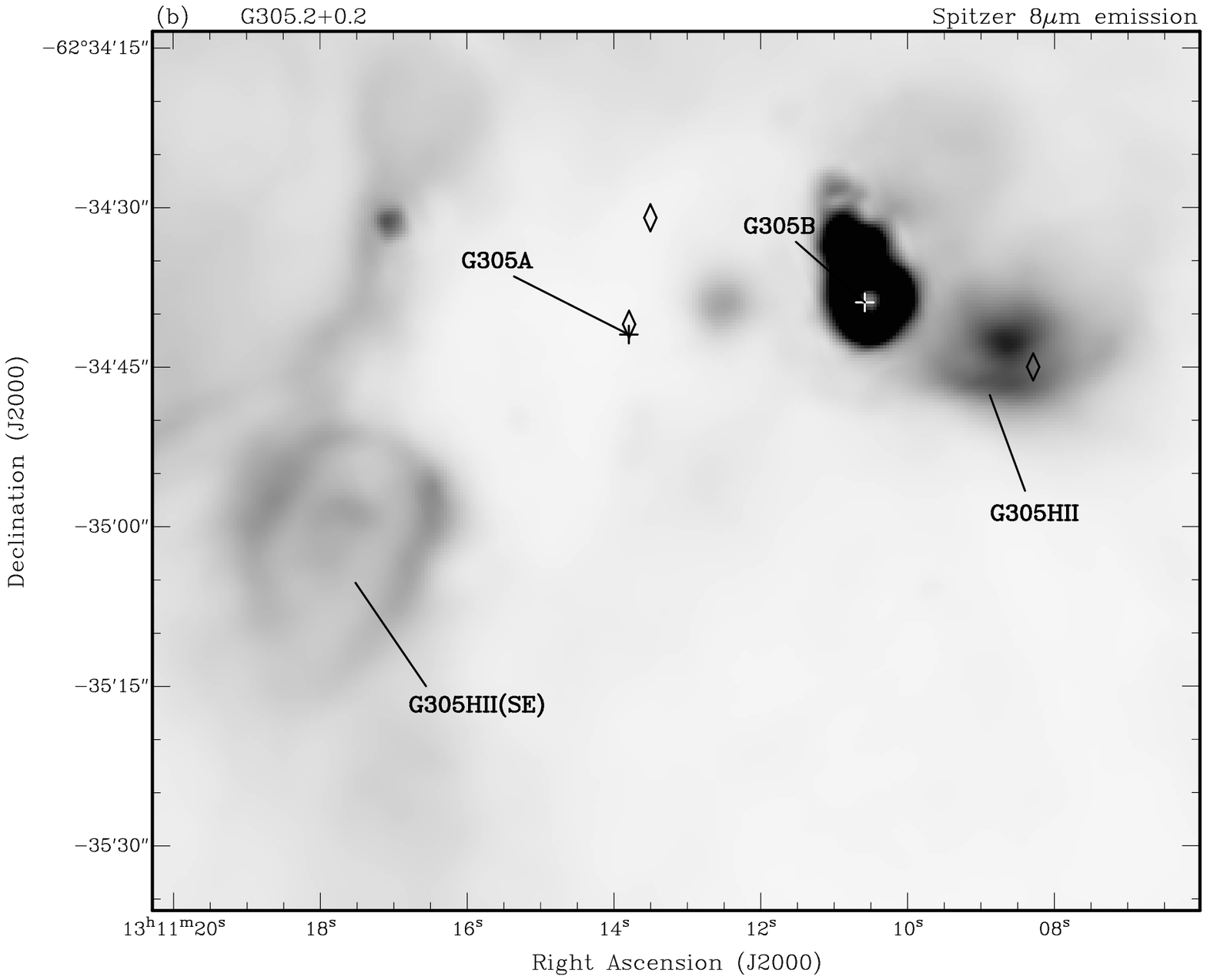}
\caption{{\bf (a)} 18496\,MHz continuum emission towards G305. The two 6.67\,GHz methanol maser sites G305A and G305B
are indicated with the plus symbols. Water masers are indicated with diamond symbols.
The contours are at 10, 20, ... 90\% of the peak flux density, which is 22\,mJy/beam. {\bf (b)} Spitzer
8$\mu$m image towards G305. Very strong emission is seen concident with G305B. This
creates an artifact that extends to the NE of G305B. No emission is seen coincident with G305A.}
\label{contimage}
\end{figure}

\begin{figure}
\includegraphics[width=1.0\columnwidth]{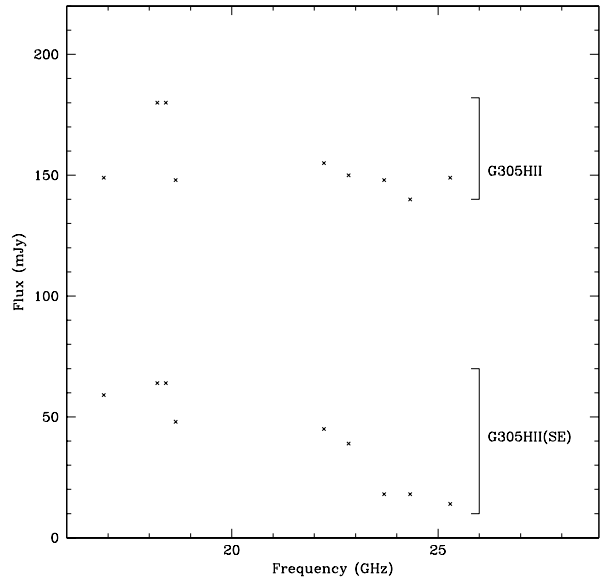}
\caption{Continuum flux as a function of frequency for G305HII and G305HII(SE). The general trend is a flat
distribution, consistent with optically thin free-free emission. The scatter seen in the points is
likely due to the errors in the absolute flux calibration for each frequency and is therefore a good representation
of the error in the flux of each \hii region.}
\label{contflux}
\end{figure}

Our 2005 May data included a frequency (18496\,MHz) devoted to continuum emission, with a bandwidth of 128\,MHz.
The continuum map is shown in Figure \ref{contimage}. Two prominent continuum features appear in the map: the
compact \hii region G305HII \citep{walsh06}, and a similarly sized compact \hii region about 30\arcsec~to
the SE of G305A, which we call G305HII(SE). We assume that each \hii region has its own powering source that is unrelated
to either G305A or G305B. This is because there is a large separation between the \hii regions and G305A or G305B.
Also, the powering source for G305HII appears as a near-infrared source at the centre of the radio continuum contours
\citep{walsh06}. The integrated intensity of the emission from G305HII is
174\,mJy and the integrated intensity of the emission from G305HII(SE) is 57\,mJy. Line-free channels from the
2005 September data can be used to reconstruct the fluxes of the \hii regions between 16 and 25\,GHz. A plot of
the distribution of fluxes with frequency is shown in Figure \ref{contflux}. Overall, it appears that the
flux remains constant over the frequency range, which is consistent with optically thin free-free emission.
There are apparent changes in flux in both sources which are matched in frequency. This is most likely due to errors
in the absolute flux calibration, which precludes us from assessing small changes in the slope of
the continuum emission. \citet{phillips98} have previously measured the continuum flux of G305HII at 8.6\,GHz
to be 195\,mJy, which is also consistent with optically thin free-free emission.

We can make an estimate of the type of stars that give rise to these \hii regions, using equation (4) of
\citet{kurtz94}. Here, we assume the distance to G305 is 3.9\,kpc, the electron temperature (T$_{\rm e}$) is
10000\,K and a($\nu$, T$_{\rm e}$) = 1. If we also assume that no photons are absorbed by dust within the
\hii regions, then we find the rate of emission of lyman continuum photons to be $2.7 \times 10^{47} {\rm s}^{-1}$
and $8.8 \times 10^{46} {\rm s}^{-1}$ for G305HII and G305HII(SE), respectively. Using the tables of
\citet{panagia73}, these rates are equivalent to slightly earlier than a B0 spectral type for G305HII
and slightly later than a B0 spectral type for G305HII(SE). If, however, 90\% of the lyman photons are absorbed
by dust within the \hii regions, then these spectral types will be revised to O8 for G305HII and O9.5 for
G305HII(SE). Note that we use the value of 90\% absorption to illustrate how this affects the spectral type. We do not
consider 90\% as the most likely figure dust over, say 10\% or 99\%.

We do not see any continuum emission associated with either of the methanol maser sites G305A or G305B, with a
3$\sigma$ detection limit of 0.15\,mJy and 0.09\,mJy, respectively. This is suprising as both sites are expected
to harbour massive stars, which should have detectable radio continuum emission, as is the case for G305HII and
G305HII(SE). G305A in particular appears to be the concentration for strong millimetre line emission (Walsh et al.
2006, and this work). This leads us to the conclusion that G305A at least, and probably G305B, are at a very
early stage of evolution where any \hii region has not had enough time to form and/or expand to an observable
size and strength.

\section{Discussion}
\subsection{Maser emission in G305.2+0.2}
\label{maser}
\begin{figure}
\includegraphics[width=1.0\columnwidth]{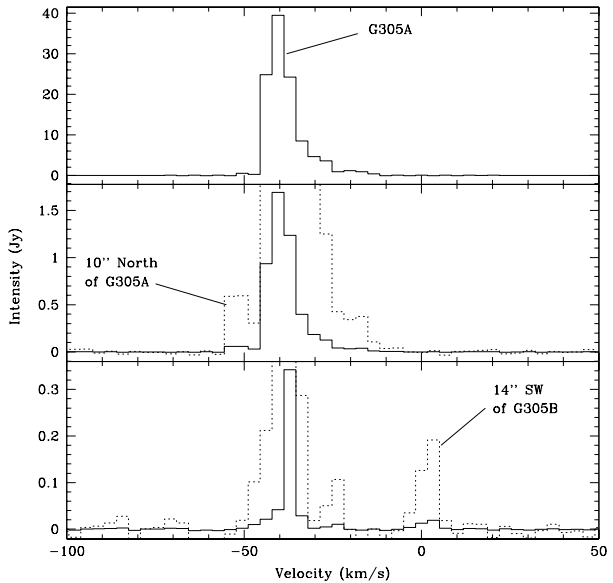}
\caption{Water maser spectra for the three detected features. The strongest feature is shown in the top spectrum
and is coincident with G305A. This feature also dominates the emission seen in the other spectra. The dotted lines
in the middle and bottom spectra show a 10-fold increase in the intensity axis, to highlight the two
weaker maser features that appear 10\arcsec~to the north of G305A (middle) and 14\arcsec~to the SW of G305B (bottom).}
\label{waterspec}
\end{figure}

As well as the two Class II 6.7\,GHz methanol maser sites G305A and G305B, a number of emission features
detected in this work may also be masers. Water masers are detected at three positions
in the field: coincident with G305A, approximately 10\arcsec~N of G305A and approximately 14\arcsec~SW
of G305B. This last maser site is coincident with the extended radio continuum emission of G305HII
(Figure \ref{contimage}).
The three water maser features appear over a velocity range of about 56\,\kms (Figure \ref{waterspec}). It is
therefore likely that at least one of these features arises in an outflow, as water masers are commonly
found in outflows (eg. \citet{moscadelli05}), which can give rise to such high velocities.

The J=4 and J=8 methanol features exhibit very bright and narrow emission about 3\arcsec~to the east of
G305B, where the Class II 6.7\,GHz methanol maser lies.
We suspect that both are Class I methanol masers \citep{mueller04}. It is suprising that
we see Class I and II methanol masers in close proximity to each other as it is unlikely that both
Class I and II masers can be strong in the same place \citep{voronkov05}. However, we note that the
3\arcsec~projected offset between the two corresponds to 12000\,AU (assuming a distance of 3.9\,kpc), which is
certainly well separated enough that there may be different physical conditions at the positions
of the Class I and II masers. Figure \ref{methimg} shows that G305A(SE) is a narrow line
emission feature in the J=4 methanol line, which may also be a maser. Due to its intrinsic weakness, we
cannot rule out the possibility that it is not a maser, but unusually narrow-lined thermal emission.
We explore the possibility that this is thermal emission in the next subsection.

\subsection{Thermal CH$_3$OH (methanol)}
\label{thermalch3oh}
\begin{figure}
\includegraphics[width=1.0\columnwidth]{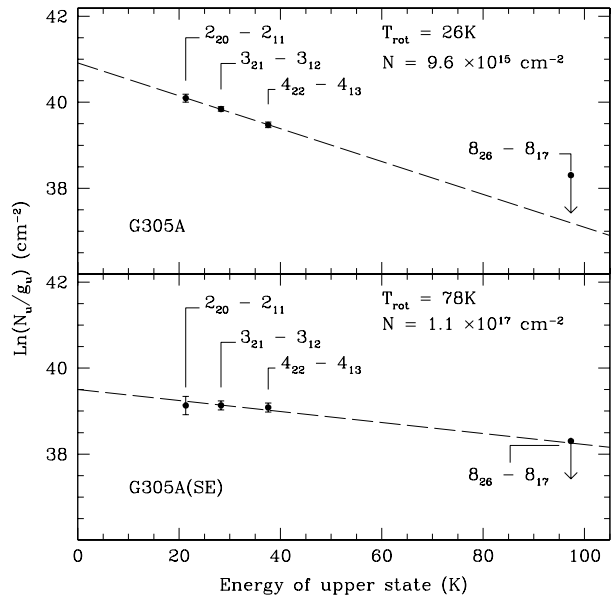}
\caption{Methanol rotation diagram for G305A (top) and G305A(SE) (bottom). For each source, a straight (dashed)
line is fitted through the three lower temperature transitions (J=2,3 and 4). The fitted rotational
temperature and column density is given for each line. Whilst the J=8 upper limit is consistent with
the fitted line for G305A, it is not possible to fit a line that is consistent with the J=8 upper limit
for G305A(SE). The dotted line in the lower plot shows a fit through the J=8 line, which results in a
column density of 1.1 $\times 10^{17}$\,cm$^{-2}$ and rotational temperature of 78\,K.}
\label{ch3ohrot}
\end{figure}

For G305A, integrated intensities of the three thermal methanol lines are listed in Table \ref{nondets}.
Assuming the lines are optically thin, we can estimate the methanol column density and rotation temperature
using a rotation diagram analysis \citep{linke79}, shown in Figure \ref{ch3ohrot}. We find a column density
of $(9.6 \pm 4) \times 10^{15} {\rm cm}^{-2}$ and temperature of 26\,K. The errorbars shown in Figure
\ref{ch3ohrot}
are determined by the relative uncertainty in the line integrated intensities. This uncertainty derives from
the fit performed to each line, as well as the inherent uncertainty in the image reconstruction process.
However, it does not include an absolute calibration error. We estimate that the error in our flux
determinations is approximately a factor of two, based on the flux calibration of the telescope.
The size of this error is confirmed by comparing 
observations of the J=3 and 4 transitions of methanol performed during both sessions in May and September.
In addition to the three detected methanol lines, we also observed the $8_{26} - 8_{17}$ transition of methanol, but
failed to detect it. Therefore, we include the upper limit for this transition in Figure \ref{ch3ohrot}.
We find that the upper limit is consistent the best fit line to the other three points.
Because the emission comes from regions which appear unresolved, we do not consider missing flux
due to missing short UV spacings a significant problem.

For G305A(SE), we also fit the three methanol lines in Figure \ref{ch3ohrot}, as well as the non-detection
in the J=8 line. Here we find a column density
of $1.1 \times 10^{17} {\rm cm}^{-2}$ and a temperature of 78\,K. 
The derived excitation temperature from the rotation diagram analysis does not have a physical
interpretation as a kinetic temperature unless the gas density is above critical.
Hence we perform, in \S\ref{sectioncoldens}, a non-LTE analysis to derive column densities and
relative abundances. As mentioned in \S\ref{maser}, the narrow
linewidths of the detected methanol transitions at this position are suggestive of maser emission.
In addition to this, the methanol spectrum at G305A(SE) is similar to the methanol maser spectrum
3\arcsec~to the east of G305B -- the J=4 feature is the strongest.
Such maser emission could explain the unusual results of the rotational diagram analysis for G305(SE) and so
we favour the explanation that this is weak maser emission.

\subsection{HC$_3$N (Cyanoacetylene)}
\label{hc3n}
Cyanoacetylene possesses hyperfine structure in its (2--1) transition.
However, no hyperfine components
were observed, apart from the main two transitions at 18196.279\,MHz. Figure \ref{hc3nimg} shows a fit to the data,
assuming a total optical depth of 0.1. The satellite components do not appear above the modelled spectrum.
Therefore, we can be confident the main line cyanoacetylene emission has an optical depth less than 0.1
(ie. optically thin).

Given the integrated intensity listed in Figure \ref{nondets}, and assuming an excitation temperature of 26\,K from the
methanol measurements,
we calculate the cyanoacetylene column density is $4.0 \times 10^{14} {\rm cm}^{-2}$.

\subsection{NH$_3$ (Ammonia)}
Ammonia is detected towards G305A as an unresolved source. Because our low spectral resolution data cannot
resolve the hyperfine structure of the (1,1) and (2,2) inversion transitions, we are unable to use our data
to investigate the optical depth of the ammonia emission. However, higher spectral and spatial resolution
ammonia observations (Steven Longmore, private communication 2007) detect (1,1), (2,2) and (4,4) emission
coincident with G305A. The (1,1) and (2,2) transitions appear optically thick, so it is not possible
to derive a reliable ammonia column density. The presence of (4,4) emission, which has an upper
energy level of 200\,K above ground, strongly suggests a hot component to G305A. 

\subsection{OCS (Carbonyl sulfide)}
\label{ocs}
Carbonyl sulfide is detected as an unresolved source towards G305A. Assuming an excitation temperature of 26\,K, we
calculate a column density of $2.5 \times 10^{15} {\rm cm}^{-2}$ for this molecule.

\subsection{H91$\beta$ radio recombination line}
This radio recombination line is detected in the field and appears coincident with the \hii region G305HII.


\subsection{G305SW - a possible prestellar massive core}
\citet{walsh06} reported the detection of G305SW in N$_2$H$^+$, but found it lacking in detectable emission in
other tracers. They suggested this may be the signature of an extremely young and cold core, potentially prestellar.
Unfortunately, G305SW lies just outside the primary beam of our 2005 May observations, but lies within the primary
beam for the 2005 September observations. We found no evidence for emission in any of the tracers from the 2005
September observations. This is not suprising for most of the more complex molecules, but it is suprising
that we did not detect any ammonia associated with G305SW. Since N$_2$H$^+$ and ammonia are chemically very similar,
we would expect them to trace similar regions.

\subsection{Derivation of column densities}\label{sectioncoldens}
The column densities listed above were derived with the
assumption that the gas is in local thermodynamic equilibrium (LTE).
The rotation diagram method used (see for example Figure 5) produces
inaccurate temperatures and column densities if only limited line
data is available ($<$3 lines). Moreover, the assumption of LTE is
only valid for large densities. We present below, column densities
derived from a non-LTE analysis of the molecular lines observed in
Table 2.

Observations have shown CH$_3$OH accounts for a significant fraction
of ice mantles relative to H$_2$O \citep{TA87}. It is believed that
CH$_3$OH is formed on the surfaces of grains by the hydrogenation of
CO on the grain surface \citep{CTM92}. Above temperatures of $\sim
90-100$\,K, all ice species, including CH$_3$OH will be sublimated
from the surfaces of grains. The abundance of CH$_3$OH should
therefore increase in the inner warmer regions of hot cores. In
\S 4.2, three CH$_3$OH lines were analyzed with a rotation
diagram and a kinetic temperature of 26\,K was derived. However, to
release the methanol from the surfaces of grains, the temperature
must have been of the order $90-100$\,K at some earlier time. It seems
unlikely that the gas would have rapidly cooled to 26K, and we
attribute the low $26$\,K temperature to the inaccurate assumption of
LTE.

In order to account for the discrepancy in the derived  low (26K)
temperature using the rotation diagram analysis of the CH$_3$OH
lines, we present below the results from a non-LTE radiative
transfer model. A new analysis of the molecular lines in Table 2
produces new column density values and temperatures, which agree
better with our understanding of the underlying physics and
chemistry (see also \S \ref{chemistrysection}). For this
calculation,  molecular data from  \citet{UMIST} is used  which
includes energy levels, statistical weights, transition frequencies
as well as collisional  rate coefficients (for collisions with
H$_2$), for each species of interest. In particular, we use
CH$_3$OH rates from \citet{pottage04}, NH$_3$ rates from
\citet{danby88}, OCS rates from \citet{green78} and \citet{UMIST}
and HC$_3$N rates from \citet{green78}. The calculation is similar to
that of \citet{UMIST}, and calculates the strength of the emission
line for a species and transition of interest, for a given density,
temperature and column density.

To derive a column density for a given species, the integrated line
intensity, $\int T_R^* dv$ in K.km/s, is found as a function of both
density (n$_{\mathrm{H_2}}$) and column density. The source is
assumed to fill the beam.
 Initially we assume some temperature and the line width
for each line of interest is taken from the observations in Table 2.
The constant contour of the observed integrated line intensity can
then be traced. This will usually give an upper or lower limit to
the column density. With more than one line, it is often possible to
simultaneously derive both a column density and density, for a given
temperature.

\begin{figure}
\includegraphics[width=85mm]{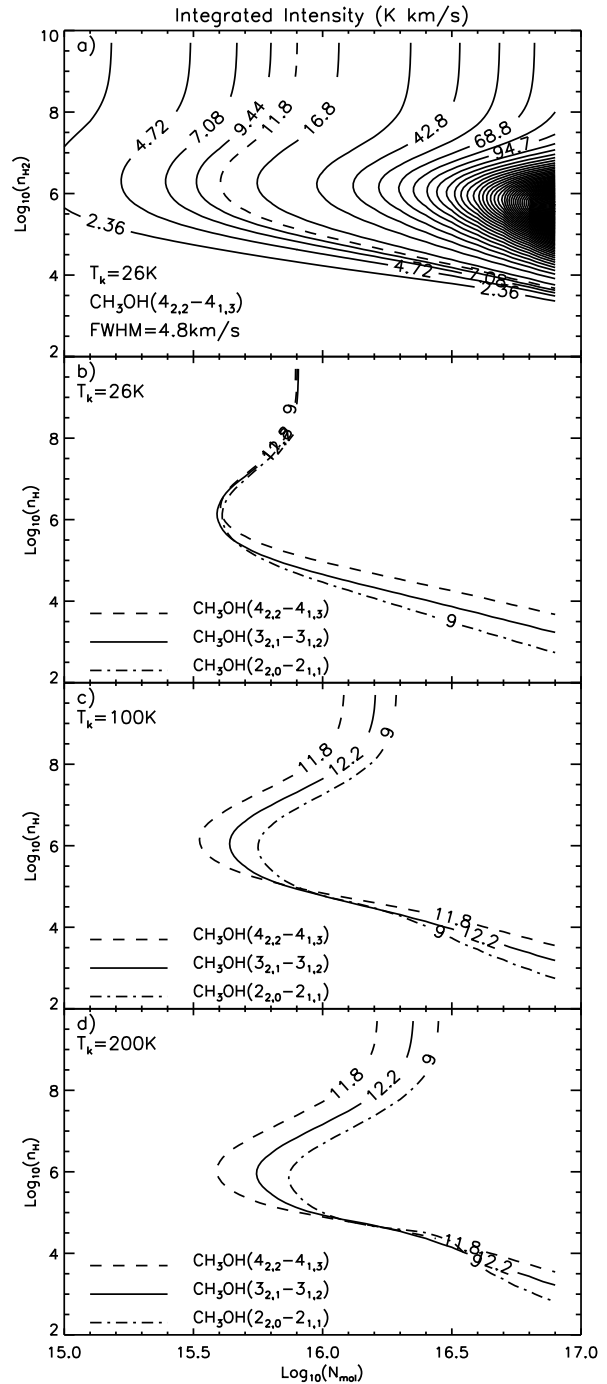}
 \caption{Contours of integrated intensity for rotational methanol
transitions. (a) Integrated intensity of the  CH$_3$OH
4$_{2,2}$--4$_{1,3}$ transition calculated using a non-LTE theoretical
model assuming a temperature of T=26\,K. The constant contour which
describes the observed intensity from Table \ref{nondets} is plotted as the
dashed line. (b) Comparison of the CH$_3$OH 4$_{2,2}$--4$_{1,3}$ observed
intensity from panel (a) with the CH$_3$OH 2$_{2,0}$--2$_{1,1}$ and
3$_{2,1}$--3$_{1,2}$ transitions calculated in a similar way for T=26\,K
using the values from Table \ref{nondets}. (c) and (d) are similar to panel (b),
but for temperatures of 100\,K and 200\,K, respectively.}
\label{figure1-CH3OHlines}
\end{figure}

Figure \ref{figure1-CH3OHlines}a shows a plot of the integrated
intensity as a function of $n_\mathrm{H_2}$  and column density
assuming a temperature of 26\,K for the CH$_3$OH($4_{2,2}-4_{1,3}$)E
line from Table 2, with a FWHM of 4.8\,km\,s$^{-1}$. The  observed
11.8\,K\,km\,s$^{-1}$
contour is plotted as a dashed line. As the density is increased and
LTE approached, the intensity becomes constant with column density.
\citet{hill05} report a density of the order $n_\mathrm{H_2}
\sim1-2\times10^4\mathrm{cm}^{-3}$, implying a column density of
$\sim 4\times10^{16}\mathrm{cm}^{-2}$ from Figure
\ref{figure1-CH3OHlines}a. However, the two other lines observed for
CH$_3$OH can be used to further restrict the column density.

In Figure \ref{figure1-CH3OHlines}b, the corresponding constant
contours of each observed integrated line intensity for all three
CH$_3$OH lines are over-plotted. All three lines overlap and agree
for $n_{\mathrm{H_2}}>10^6$cm$^{-3}$, yielding the LTE limit of the
column density. Since the observations of \citet{hill05} suggest a
lower density, it is likely that the gas is not in LTE. In Figures
\ref{figure1-CH3OHlines}c and \ref{figure1-CH3OHlines}d, the same
lines are over-plotted assuming temperatures of 100K and 200K.
In each of these cases, there is agreeement for densities and column
densities in the ranges  $ 1.8\times 10^4$cm$^{-3} < n_{\mathrm{H_2}}
< 8 \times 10^4$cm$^{-3}$ and $7.9\times 10^{15}$cm$^{-2}$$ <N < 2.0
\times 10^{16}$cm$^{-2}$ respectively. Since all three lines agree
for both temperatures, this method does not constrain the
temperature. However, as previously mentioned, temperatures
of 90\,K or more are expected to evaporate CH$_3$OH off grain surfaces,
and so we favour a temperature of at least 100\,K.

\begin{figure}
\includegraphics[width=85mm]{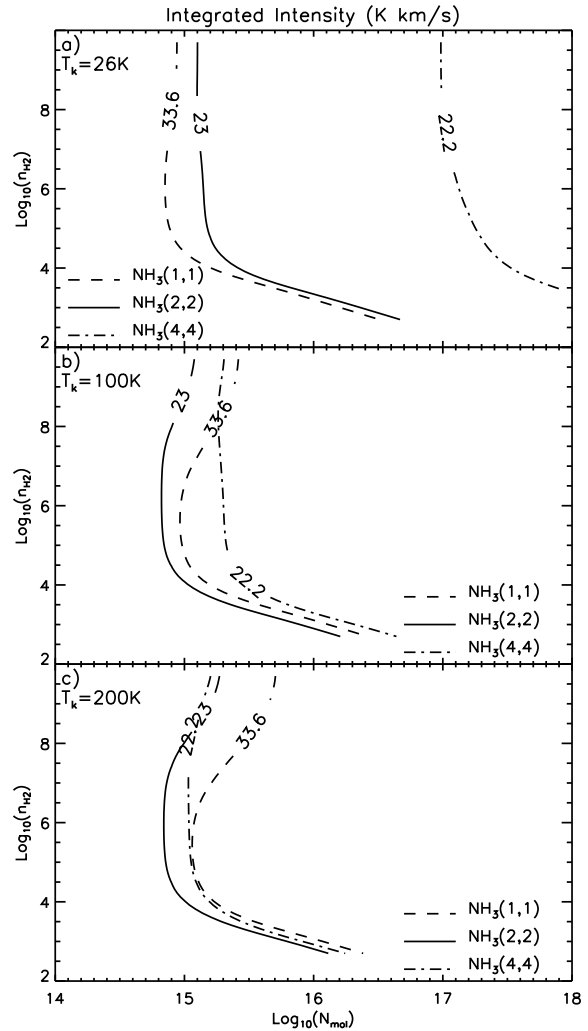}
 \caption{Contours of integrated intensity for rotational ammonia
transitions. (a) Integrated intensities of the NH$_3$ rotation
transitions (1,1), (2,2) and (4,4) calculated using a non-LTE
theoretical model assuming a temperature of T=26\,K. The chosen
contours are taken from the observed intensities of Table \ref{longmoretable}.
In panels (b) and (c) the same comparisons are made, assuming
temperatures of 100\,K and 200\,K, respectively.}
\label{figure2-NH3lines}
\end{figure}

High spectral and spatial resolution ammonia observations have been
carried out (Steven Longmore, private communication 2007) for the (1,1),
(2,2) and (4,4) lines. The details of the detection of these
lines coincident with G305A are given in Table
\ref{longmoretable}. A similar analysis to that of Figure
\ref{figure1-CH3OHlines} for the three NH$_3$ lines is shown in
Figure \ref{figure2-NH3lines}. In Figure  \ref{figure2-NH3lines}a,
the temperature is 26\,K, and clearly there is no agreement between the
NH$_3$(4,4) line and the other two lines. However, if the
temperature is increased, as  in Figures \ref{figure2-NH3lines}b and
\ref{figure2-NH3lines}c then there is closer agreement. The value
derived from these plots for the column density is a little
ambiguous since the (1,1) and (2,2) contours only converge as
n$_{\mathrm{H_2}}$ decreases, so we only quote a lower limit of
$N_{\mathrm{NH_3}}>10^{15}$cm$^{-2}$. However, it is important to
note that the observation of the NH$_3$(4,4) strongly suggests that
there is a temperature in excess of 100\,K.

We note that densities of $1.8\times10^4-8\times10^4{\rm cm}^{-3}$ are
unusually low for traditional hot core densities ($\sim10^6{\rm cm}^{-3}$).
However, complex organic molecules are seen in extended emission in SgrB2 and  
in the central molecular zone \citep{requena06}. \citet{requena06} find that these
molecules, including CH$_3$OH, can have low excitation temperatures (10-20\,K) despite high
kinetic temperatures ($\sim$100\,K) and cold dust ($<$20\,K) because the density is
low ($\sim10^4{\rm cm}^{-3}$). They argue that shocks remove mantle material and drive
`hot core' type chemistry. So low excitation temperatures may not
necessarily reflect cold gas-phase chemistry.

\begin{table}
 \centering
  \caption{Observations G305A from Steven Longmore (private communication 2007).}\label{longmoretable}
  \begin{tabular}{lllll}
  \hline
Observed line & Integrated Intensity  & Line Width  \\
              & (K\,km\,s$^{-1}$)         & (km\,s$^{-1}$)  \\
    \hline
NH$_3$(1,1)   &     33.58            &           5.5   \\
NH$_3$(2,2)   & 23.05                &     5.1      \\
 NH$_3$(4,4) &   22.23               &     7.9    \\
    \hline
\end{tabular}
\end{table}

\begin{figure}
\includegraphics[width=85mm]{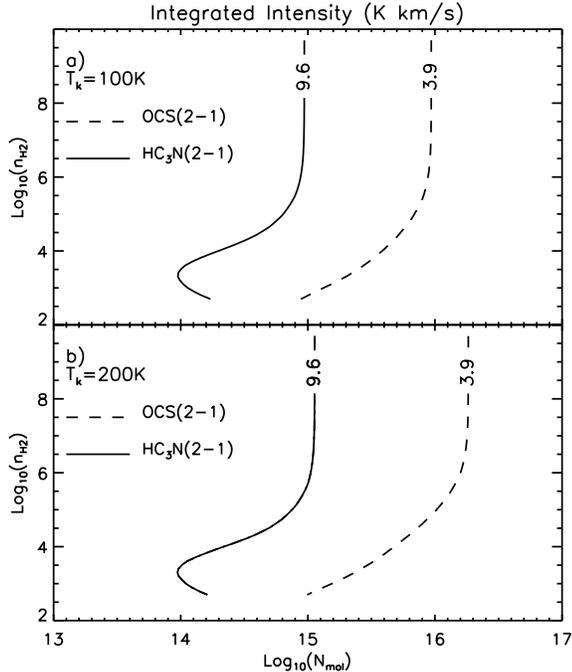}
 \caption{Contours of integrated intensity for rotational OCS and
HC$_3$N transitions. (a) Integrated intensities of the OCS (2--1) and
HC$_3$N rotation transitions calculated using a non-LTE theoretical
model  assuming a  temperature of T=100\,K. The chosen contours are
taken from the observed intensities of Table \ref{nondets}. (b) the same
transitions are presented as in panel (a), but with a temperature of
200\,K.}
\label{figure3-HC3N-OCSlines}
\end{figure}
We cannot do  a complete analysis of OCS and HC$_3$N, with only
single transitions for each. However we can give new values assuming
densities in the range $ 1.8\times 10^4$cm$^{-3} < n_{\mathrm{H_2}} <
8 \times 10^4$cm$^{-3}$.  Figure \ref{figure3-HC3N-OCSlines} shows
the OCS(2--1) and HC$_3$N(2--1) lines plotted for 100K and 200K. For
densities in the range $ 1.8\times 10^4$cm$^{-3} < n_{\mathrm{H_2}} <
8 \times 10^4$cm$^{-3}$, we can put limits on the column densities,
which are given in Table \ref{nonLTEcoldens} for 100K, along with
the other non-LTE values for NH$_3$ and CH$_3$OH. We note that column
densities derived here compare well to those derived in sections \ref{thermalch3oh},
\ref{hc3n} and \ref{ocs}, which assume LTE. This indicates that column
densities derived under the assumption of LTE, at least in the case of G305A,
may provide reasonable estimates of the true column density.

\begin{table}
 \centering
  \caption{Column densities for G305A from \S
  \ref{sectioncoldens}, for $ 1.8\times 10^4$cm$^{-3} < n_{\mathrm{H_2}} < 8 \times
10^4$cm$^{-3}$.}
\label{nonLTEcoldens}
  \begin{tabular}{lll}
  \hline
Molecule & Column Density (cm$^{-2}$)    \\
    \hline
CH$_3$OH   & $7.9\times 10^{15}$cm$^{-2}$$ <N < 2.0 \times
10^{16}$cm$^{-2}$    \\
NH$_3^*$  &  $>10^{15}$ \\
OCS(100K)      &   $4.0\times 10^{15}<N_{\mathrm{OCS}} <6.3 \times  10^{15}$  \\
OCS(200K)      & $5.6 \times 10^{15}<N_{\mathrm{OCS}} < 9.5 \times 10^{15} $    \\
HC$_3$N(100K)  &  $2.0 \times 10^{14}<N_{\mathrm{HC_3N}} < 6.3 \times 10 ^{14}$ \\
HC$_3$N(200K)  &  $2.8 \times 10^{14} <N_{\mathrm{HC_3N}} < 6.3 \times 10^{14}$   \\
    \hline
\end{tabular}
$^*$value derived from observations of Steven Longmore (private communication, 2007).
\label{revised}
\end{table}

\begin{figure}
\includegraphics[width=85mm]{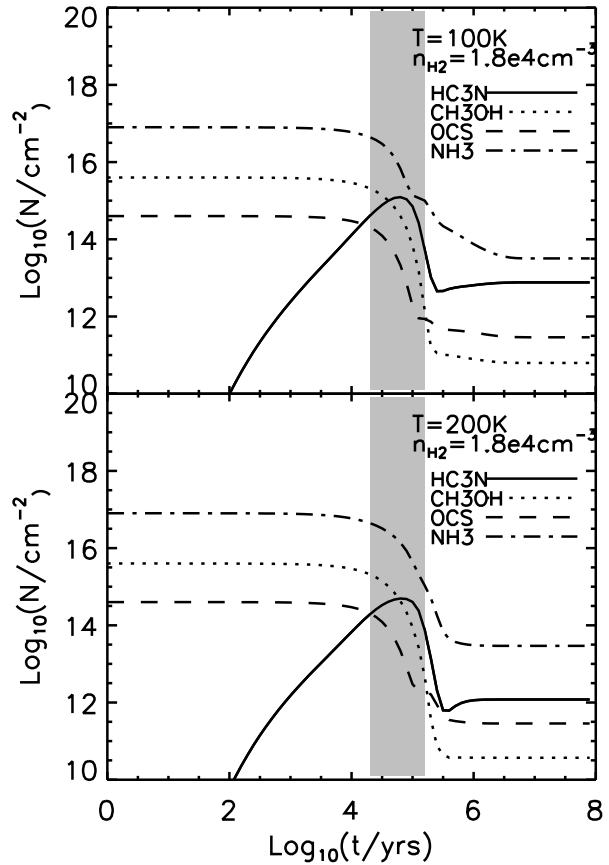}
 \caption{Time dependent column densities for the four species of
interest, as calculated from a spherical hot core chemical model of
G305A as described in \S\ref{chemistrysection}. A constant density is
assumed, $n_{\mathrm{H_2}}=1.8\times10^4$cm$^{-3}$, with a
temperature of 100\,K (top panel) or 200\,K ( bottom panel). A
comparison of a species column density constrains the age of the
core. The grey shaded bar illustrates the age range of the core
derived in comparing the observed column densities given in Table
\ref{revised}.}
\label{figure4-model_coldens}
\end{figure}

\begin{figure}
\includegraphics[width=85mm]{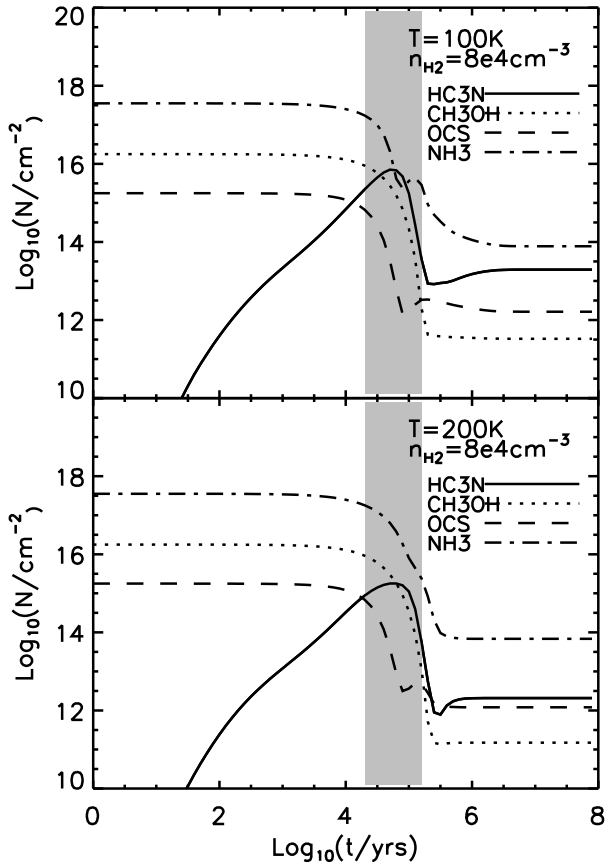}
 \caption{Column densities from the spherical core model, similar to
Figure \ref{figure4-model_coldens}, but with the density increased to
$n_{\mathrm{H_2}}=8\times10^4$cm$^{-3}$.}
\label{figure5-model_coldens}
\end{figure}

\section{Time-Dependent Chemistry}\label{chemistrysection}

The analysis in section 4, in particular the detection of the NH$_3$ (4,4)
line, strongly suggest that G305A has a hot component with a temperature
of the order 100 -- 200\,K. We will therefore now model the chemistry of
G305A, based on the assumption that it is a hot core. The time dependent
chemistry of a spherical core, representative of G305A, will be calculated
below and the observations will be discussed in light of these results.
A molecule found inside a hot core may be present
due the following three processes: if it is either formed by cold
chemistry in the earlier core collapse phase then accreted onto a
grain and re-released inside the hot core, formed by grain surface
chemistry and released, or formed in the gas phase chemistry of the
hot core. Although the mantle compositions are relatively simple
e.g. CH$_3$OH and NH$_3$, the release of these into the gas phase
are believed to be the drivers of the very complex chemistry
observed in hot cores. The relative abundances of the mantle
species, along with varying densities and temperatures can lead to
chemical differences among cores.

We now attempt to model the gas phase chemistry of G305A. We assume
the core is spherical with a density $n_{\mathrm{H_2}}=1.8 \times
10^4 $cm$^{-3}$ and  $n_{\mathrm{H_2}}=8 \times 10^4 $cm$^{-3}$ based
on the analysis in section \ref{sectioncoldens}, and a radius of
1.1\,pc \citep{hill05}. The model is run twice with constant
temperatures throughout  of  100\,K and 200\,K.  We use a model similar
to \citet{Mea97}, which consists of 22 depth points, or shells, each
with a specified density and temperature. Column densities are
calculated by integrating along lines of sight, and to account for
beam dilution, the column densities are weighted and summed over a
gaussian beam (eg., \citealt{TMM99}). The source is assumed to fill
the beam. 211 species are chosen
with 2575 reactions and rate coefficients, taken from the online
UMIST database \citep{leteuff00}.

The initial abundances are taken from Table 2 of \citet{Mea97} which
assumes that the gas is mostly molecular with some ionization by
cosmic rays. Assuming spontaneous evaporation from grain surfaces,
there are also the primary injected mantle molecules CH$_3$OH,
NH$_3$, C$_2$H$_2$,C$_2$H$_4$,C$_2$H$_6$,CH$_4$, H$_2$CO, CO$_2$,
O$_2$, H$_2$O, H$_2$S. Ice absorbtion spectra towards massive young
stellar objects suggest that OCS should also be included as a mantle
species \citep{kea01}. We also find that an initial abundance of OCS
is needed in the model to generate an OCS abundance that resembles
the observations, the gas phase chemistry alone cannot produce the
observed column density (see the discussion for Figures
\ref{figure4-model_coldens} and \ref{figure5-model_coldens} below).
The initial abundances are given in Table
\ref{tableinitialabundance} which are based on the models of
\citet{Mea97} and \cite{NM04}. The initial conditions are also chosen
so that the total abundance of C, N and O do not exceed  with the
mean interstellar values; C/H $= 1.4\times10^{-4}$, O/H$=3.19\times
10^{-4}$ and N/H $= 7.5 \times 10^{-5}$ \citep{Cea96, Meyerea97,
Meyerea98}. It should  be noted that differences in the mantle
chemistry and choice of initial abundances  could drive a different
gas phase chemistry, this will be considered in future papers, and
is outside the scope of the current work.

\begin{table}
 \centering
 \begin{minipage}{90mm}
  \caption{The initial abundances, with respect to H nuclei, assumed in the chemical model }\label{tableinitialabundance}
  \begin{tabular}{llll}
  \hline
Species   & Abundance & Species   & Abundance   \\
\hline \\
CH$_3$OH & $5\times10^{-7}$ & C$_2$H$_2$  & $5\times 10^{-7}$   \\
NH$_3$ &  $1 \times 10^{-5}$ & C$_2$H$_4$ & $5 \times 10^{-9}$    \\
H$_2$S & $1 \times 10 ^{-6}$& C$_2$H$_6$  & $5 \times 10^{-9}$    \\
H$_2$O & $1 \times 10 ^{-5}$& CH$_4$      & $2 \times 10^{-7}$   \\
O$_2$  & $1 \times 10^{-6}$ & CO$_2$      & $5.0 \times 10^{-6}$  \\
H$_2$CO & $4 \times 10 ^{-8}$& OCS        & $5.0 \times 10^{-8}$ \\
CO     & $5 \times 10^{-5} $&  N$_2$      & $2 \times 10^{-7}$  \\
He$^+$     & $2.5 \times 10^{-11}$ & H$^+$& $1 \times 10^{-10} $ \\
H$_3^+$    &$ 1 \times 10^{-8}$ &  Si$^+$ & $3.6 \times 10^{-8}$ \\
Fe$^+$       & $ 2.4 \times 10^{-8}$&\\
 \hline
\end{tabular}
\end{minipage}
\end{table}

The model results are presented in Figures \ref{figure4-model_coldens}
and \ref{figure5-model_coldens} for
$n_{\mathrm{H_2}}=1.8\times10^4$cm$^{-3}$ and
$n_{\mathrm{H_2}}=8\times10^4$cm$^{-3}$, respectively. The column
densities of the four species of interest are plotted as a function
of time since the onset of the core heating. In each figure the
chemical model has been run twice assuming temperatures of 100\,K
and 200\,K. By comparing the column densities for each species in
Table \ref{revised}, derived from the observed emission lines in Table
\ref{nondets}, the
chemical model can be used to constrain both the temperature and age
of the core.
In the gas phase, the production of  CH$_3$OH is very inefficient (eg.
\citet{woodall07}); going via radiative association of CH$_3^+$ and H$_2$O
at a low rate \citep{Lucaetal02}, while \citet{Getal06} have recently shown
that the fraction of dissociative recombinations which produce methanol are of the order
3\%.
From the methanol column density limits quoted in Table
\ref{nonLTEcoldens}, upper age limits are derived in Figure
\ref{figure5-model_coldens} for $n_{\mathrm{H_2}}=8 \times 10^4
$cm$^{-3}$ ; $3.1\times 10 ^4 $  years for 100\,K and
 $2\times 10^4$ years from 200\,K.
The lower density $n_{\mathrm{H_2}}=1.8 \times 10^4 $cm$^{-3}$ cases
of Figure \ref{figure4-model_coldens} do little to constrain the age
since $N_{\mathrm{CH_3OH}}< 7.9 \times 10^{15}$cm$^{-2}$.  From the
NH$_3$ observations, both Figures \ref{figure4-model_coldens}  and
\ref{figure5-model_coldens} give upper age limits of $1.6 \times
10^5$ years.

The column density of HC$_3$N indicates either, a minimum age of
$\sim 10^4$ years, or an age of $10^5$ years. We favor the later
time based on the  NH$_3$ comparisons discussed above. As noted by
\citet{Mea97}, HC$_3$N is usually regarded as a cold cloud tracer,
however it is formed in a hot gas through reactions of N and
C$_3$H$_2$ and C$_2$H$_2$ with CN, where C$_2$H$_2$ is evaporated
from grain mantles. This suggests that HC$_3$N should be a useful
species to trace the conditions and ages of hot cores.  The highly
variable HC$_3$N  abundance with time, in Figures
\ref{figure4-model_coldens} and \ref{figure5-model_coldens}, further
supports the use of HC$_3$N as a  `chemical clock' species.

The OCS column densities  quoted in Table \ref{nonLTEcoldens} are
$>4\times 10^{15}$ cm$^{-3}$, which is higher than the maximum OCS
column density for all cases. The observed OCS abundance can be
reproduced in the model if the initial abundance of OCS was increased.
For simplicity, we assumed the OCS abundance from \cite{NM04} based
on their model of G34.3+0.15, and so it is highly possible that the
initial  OCS  abundance was  greater for G305A. We will examine the
effects of changes in the model for varying initial conditions in
future papers. It is important to note that if no initial abundance
of OCS is assumed in the model, then the gas phase chemistry only
produces a OCS column density of the order $4 \times
10^{13}$cm$^{-2}$ (plots not shown here), which strongly suggests
that the origin of the OCS is the evaporated ice mantles.
Moreover, OCS has been detected in ice-absorbtion features of massive
protostars \citep{kea01}.

It is important to note that  the column densities of NH$_3$ and
HC$_3$N give similar core ages, $\sim 10^5$years, for each of the 4
cases plotted in Figures \ref{figure4-model_coldens} and
\ref{figure5-model_coldens}. However, the determination of the
temperature and density is still a little ambiguous. There is better
agreement for NH$_3$ and HC$_3$N times for the denser hotter case.
Moreover, the NH$_3$ observations (Steven Longmore, private
communication 2007) discussed in \S\ref{sectioncoldens} ( in
particular Figure \ref{figure2-NH3lines}) suggest higher
temperatures of the order 200\,K. We therefore favor the high density
$8\times 10^4$cm$^{-3}$ and 200\,K case,  suggesting a core age in the
range $2 \times 10^4 <t_{core} <1.5 \times 10^5$ years.

\section{Conclusions}
We have observed the massive star forming region G305.2+0.2 at 1.2\,cm. We detected emission in
methanol, cyanoacetylene, ammonia, carbonyl sulfide and water, as well as the radio recombination
line H91$\beta$. All molecular transitions are detected towards G305A, which confirms the hot core nature
of this source. Three lines of methanol are found in emission at the position of G305A, and together with a
fourth transition which was not detected, we have constructed a rotational diagram where we derive a rotational
temperature of 26\,K and a column density of 9.6 $\times 10^{15}\,{\rm cm}^{-2}$ for methanol. However, because
methanol requires temperatures in excess of 90\,K to evaporate from dust grains, we find the temperature unrealistic
and attribute this to the incorrect assumption of LTE.
With a non-LTE analysis of G305A, we derive a model which is consistent with
a density of $8 \times 10^4$cm$^{-3}$, and a temperature of the order 200\,K.
A gas phase time dependent chemical model further constrains the age of the
core to $2 \times 10^4 <t_{core} <1.5 \times 10^5$ years.

Extended emission to the SE of G305A is seen in three methanol transitions. It has a narrow line width and an unusually
high rotational temperature (78\,K). We interpret this as evidence that the emission may be a weak maser.

Strong methanol maser emission is found 3\arcsec~to the east of G305B. Since the methanol transition at 24.933\,GHz
is a Class I maser, G305B appears to be a unique object worthy of further investigation as it shows both Class I and II
methanol maser emission within a small region. No other detections were made at the position of G305B. Since it is also
known to be a luminous source, the lack of molecular emission suggests G305B is an older source that has already
had time to clear out its surrounding molecular material.

Radio continuum emission at 18.496\,GHz is detected towards two \hii regions (G305HII and G305HII(SE)). G305HII shows
a cometary morphology and is probably powered by a star in the range O8 to B0, whereas G305HII(SE) shows a shell
morphology and is probably powered by a star in the range O9.5 to B0.

No radio continuum emission was detected towards either G305A or G305B. For the case of G305A, this may be
because the source is too young to have produced an observable \hii region.
However, it is not clear why the older, and infrared-luminous, source G305B does not show any
radio continuum emission.

\section*{Acknowledgments}
We would like to thank the anonymous referee who has helped improve the quality of this work.
We would like to thank Andrej Sobolev for helpful discussions.
We would also like to acknowledge the work of Steven Longmore who helped with discussions and
data of ammonia. The Australia Telescope is funded by the Commonwealth of Australia for operation as a National
Facility managed by CSIRO. This research was supported under the Australian Research Council's
Discovery funding scheme (project number DP051893). Astrophysics at QUB is supported by a
grant from PPARC.


\end{document}